\documentclass[10pt]{article}
\usepackage{color}
\usepackage{amsmath} 
\usepackage{amssymb} 
\usepackage[english]{babel}
\usepackage{graphics}
\usepackage[pdftex]{graphicx}
\usepackage{xspace}
\usepackage{gensymb}
\usepackage[gen]{eurosym}
\usepackage[font=footnotesize,format=plain,labelfont=bf,up,textfont=it,up]{caption}
\usepackage[left]{lineno}
\usepackage{sectsty}
\sectionfont{\large}
\subsectionfont{\small}

\newcommand{\nubar}{$\overline{\nu}\ $}
\newcommand{\nue}{\ensuremath{\nu_{e}}\xspace}
\newcommand{\nubare}{\ensuremath{\overline{\nu}_{e}}\xspace}

\newcommand{\numu}{\ensuremath{\nu_{\mu}}\xspace}

\newcommand{\nutau}{\ensuremath{\nu_{\tau}}\xspace}
\newcommand{\numunue}{\ensuremath{\nu_\mu \rightarrow \nu_e}\xspace}

\newcommand{\pnuenue}{\ensuremath{P(\nue \rightarrow \nue)}\xspace}
\newcommand{\pnuenumu}{\ensuremath{P(\nue \rightarrow \numu)}\xspace}

\newcommand{\nuenumu}{\ensuremath{\nue \rightarrow \numu\xspace}}

\newcommand{\nubarmunubare}{\ensuremath{\overline{\nu}_\mu \rightarrow \overline{\nu}_e}\xspace}

\newcommand{\boss}[2]{\ensuremath{\rlap{\kern-2.5pt\ensuremath{\overset{\scriptscriptstyle(-)}{\phantom{#1}}}}{\ensuremath{{#1}_{#2}}}}}

\begin{document}






\vskip 20pt
\begin{center}
{\bf\Large  A View of Neutrino Studies with the Next Generation Facilities}
\vskip 10pt
L.~Stanco

 {\em \footnotesize INFN-Padova, Via Marzolo, 8, I-35131 Padova, Italy}
\end{center}

\vskip 20pt

\vskip 20pt
\centerline{{\bf Abstract}}

\vskip 20pt

\begin{center}
\begin{minipage}{5in}
Neutrino physics is nowadays receiving more and more attention as a possible source of information for the long--standing
investigation
of new physics beyond the Standard Model. The rather recent measurement of the third mixing angle $\theta_{13}$ in the standard 
mixing oscillation scenario encourages the pursuit of what is still missing: the size of any leptonic CP violation, absolute neutrino
masses and the characteristic nature of the neutrino. 
Several projects are currently running and they are providing impressive results. 
In this review, 
the phenomenology of neutrino oscillations that results from the last two decades of investigations is reviewed,
with emphasis on our current knowledge
and on what lesson can be taken from the past. 
We then present a critical discussion of current studies on the mass ordering and what might be expected from future results. 
Our conclusion is that decisions determining the next generation of experiments and investigations have to be strictly based on the 
findings of the current generation of experiment. In this sense it would be wise to wait a few years before taking decisions on the future projects. In the meantime, since no direct path forward is evident for the future projects, the community must be 
committed to their careful evaluation.
\end{minipage}
\end{center}

\vskip 20pt

\section{Introduction}\label{sec:intro}

The current scenario of the Standard Model (SM) of particle physics, being arguably stalled by the discovery of the Higgs boson,
is {\em desperately} looking for new experimental inputs to provide a more comfortable theory.
In parallel, experiments on neutrinos so far have been  an outstanding source of novelty and unprecedented 
results. In the last two decades several results were obtained 
by studying atmospheric, solar or reactor neutrinos, or more recently with neutrino productions 
from accelerator--based beams. Almost all these results have contributed to strengthen 
the flavour--SM. 
Nevertheless, 
relevant parts like the values of the leptonic CP phase and the neutrino masses are still missing, a critical ingredient
being the still undetermined neutrino mass ordering.
On top of that the possibility of lepton flavour violation (if e.g. neutrinos are Majorana particles), 
is a very open issue, experimentally strongly pursued. 

Even if the Standard Model can be easily extended with right-handed neutrinos to introduce Dirac mass terms,
notwithstanding the lightness of the neutrino masses points to very small and unnatural Yukawa couplings. The latter issue is likely overcome
by considering a Majorana neutrino mass and some choices of see-saw mechanisms. 
This peculiarity of neutrinos, compared
to the other charged fermions, originates from the fact that they are neutral particles. The possible Majorana nature
of neutrinos would correspond to 
lepton--flavour violation and a real portal for new physics beyond the SM. 
It is intriguing and wishful that studies on
neutrinos could uncover some of the solutions to the open questions in fundamental physics. However, 
it might even happen that all our hopes are shattered in the end,
and a coherent picture of SM will continue to hold, i.e. three flavours with no Majorana mass and a 
{\em natural} mass hierarchy, together with a tiny leptonic CP--phase.

Nevertheless, there are already some measurements that do not fit the standard 3--flavour neutrino--framework hinting instead at the possible 
existence of one (or more) {\em sterile} neutrino. That is a very wide issue, as well as one experimentally strongly pursued, too. 
If the existence of the dark matter and its possible interplay to neutrinos are additionally taken into consideration,
thus the present picture turns out to be very stimulating.

From 2012 neutrino--oscillation physics entered a new era, as many applicable measurements were collected in the meantime. 
From one side, phenomenological fits were continuously improved by inputs given by those measurements. A coherent picture could be expected to emerge for the four most relevant missing pieces,
namely, the CP--phase, the mass ordering, the octant  of the largest mixing angle $\theta_{23}$, and
the presence or not of new sterile--like states at the eV mass scale. In any case  
the phenomenological scenario will be tested by the ensemble of inputs providing
either a coherent or not--coherent picture. 
From the other side, many new experimental proposals were put forward, even if some of them not yet fully funded. 
In the context of the strategy the neutrino community is requested to take for the future, all of these proposed future projects must be 
carefully evaluated and perhaps even rejected in the event that the currently running experiments and approved projects will be able to 
confirm and complete
the {\em standard} scenario by the year 2020--2025 
(or less).

It would be unconceivable even to think to include in this short review descriptions of all the facts today known about neutrinos
together with an exhaustive discussion of the whole set of experiments and proposals for the near future.
Therefore a concise attitude is adopted, either referring to the bibliography or not including on purpose many results/studies/projects 
not so relevant to the mainstream of the discussion, which is instead focussed on the major issues
according to the judgement 
of the author.
The paper is organized as follows. In the next Section an overview of the acquired phenomenological scenario for neutrinos  is presented, while in the 
following one a critical discussion on the future determination of the mass ordering/hierarchy (MH) is depicted.
A brief description of the major on--going experiments and fully funded proposals, useful to the mainstream, follows. In the last Section some final considerations and conclusions
are drawn. Several issues are just mentioned and not developed, 
as attempted measurements of individual neutrino masses,
and the studies on the production and detection of the solar and supernova neutrinos.

\section{Neutrino Phenomenology in the last two decades and nowadays}\label{sec:pheno}

The most famous {\em hunter} of neutrinos is probably Raymond Davis, Jr.. From the late sixties, with collaborators  he looked at neutrinos coming 
from the Sun~\cite{davis}. It took almost 3 decades to collect about 2,000 solar electron--neutrino candidates in the Homestake experiment, much less (about 1/3) than what predicted by John N. Bahcall and collaborators (see, e.g., Ref.~\cite{bahcall} for a discussion). Even if the neutrino deficit  w.r.t. the solar models was unveiled quite soon~\cite{bahcall-2}, the dispute
was finally  settled by the confirmation of the neutrino oscillation. That was reached by the observation of the
oscillations in both the atmospheric--neutrino sector by
Super-Kamiokande (SK) in 1998~\cite{superk} through the \numu disappearance\footnote{The correct inheritance of the physics measurements and results on 
atmospheric neutrinos is more 
articulated than here reported. More experiments were actually involved, see e.g.~\cite{gg}.} and the solar sector by SNO in 2002~\cite{sno}
through the measurement
of the neutral current (NC) interactions, equally sensitive to all the neutrino flavours. The NC measurement confirmed the  predictions of the solar model, and therefore the rightness of the deficit by Davis\&Bahcall  as due to a flavour changing of neutrinos from the Sun.

\begin{figure}[htbp]
\begin{center}
  \includegraphics[width=0.6\textwidth]{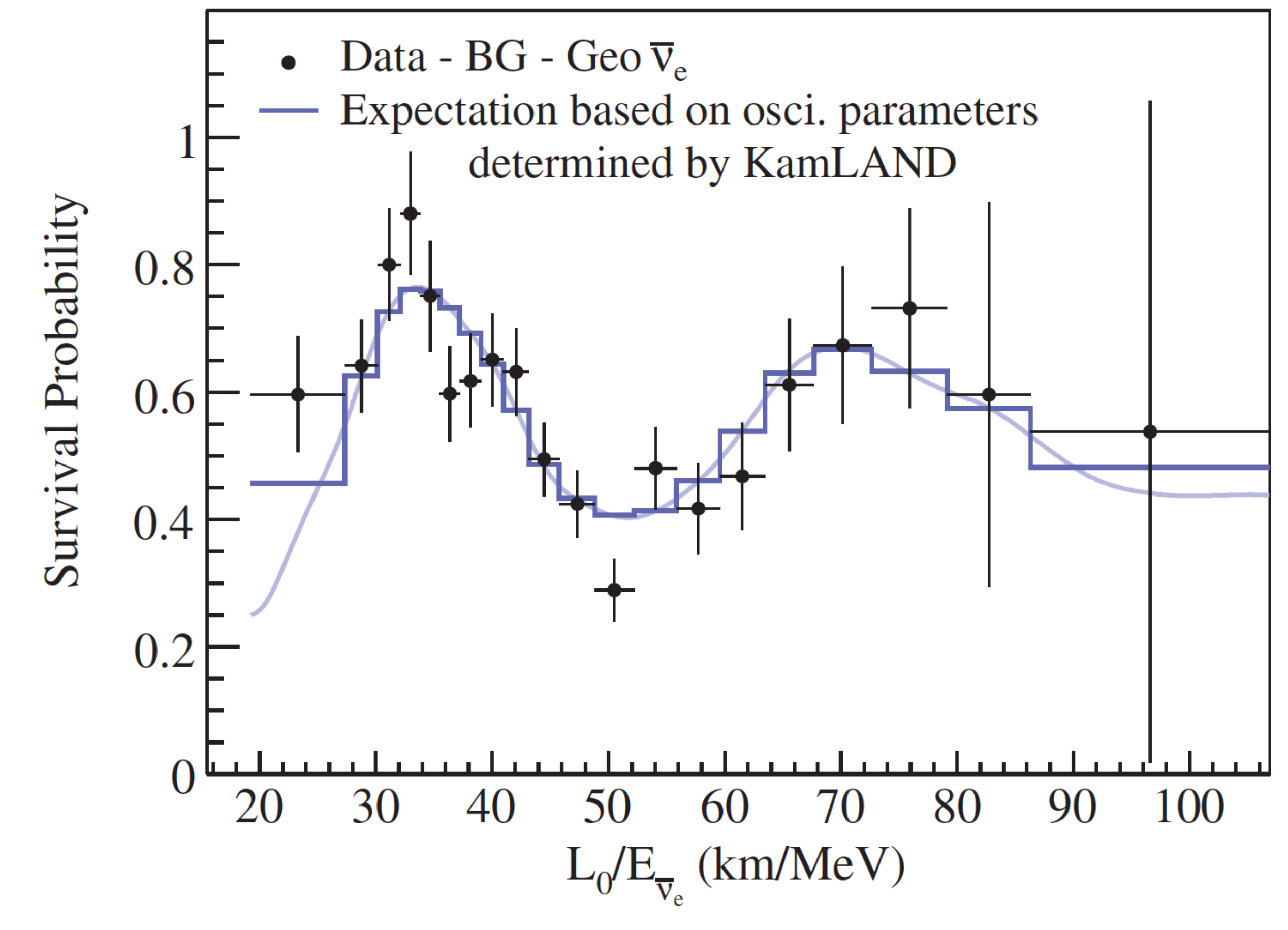}%

    \caption{The neutrino oscillation  pattern measured by KamLAND from the reactor flux (figure~3 of~\cite{kam-fig}). The histogram shows
     the ratio of the observed background--subtracted \nubare spectrum to the
     expectation for no--oscillation as a function of L$_0$/E. L$_0=180$ km is the effective baseline, as if all anti--neutrinos 
     detected in KamLAND were due to a single reactor at this distance. The energy bins are equal probability bins of the best fit including 
     all backgrounds (mainly accidentals, neutron related from $\alpha$--decay of $^{210}$Po and secondary interaction from 
     $^{13}$C($\alpha$,$n$)$^{16}$O, and geoneutrinos, attainable to a total of about 20\% of the observed 1609 events, in the ratio 0.05:0.10:0.05, respectively).
     The curve shows the expectation based on the best fit oscillation parameters.}
    \label{fig:kamland}
\end{center}
\end{figure}

However, the just evident neutrino mass mixing was again puzzled by the simultaneous null result of CHOOZ in 
1998~\cite{chooz} that looked at neutrino oscillations at a very short distance (1~km) from an anti--\nue reactor flux.
The puzzle on flavours was clarified in 2002  after the KamLAND~\cite{kamland} measurement of the reactor--neutrino flux 
at an averaged distance of 180~km from several 
nuclear power plants. KamLAND showed evidence of the spectral distortion as function
of $L/E$ (distance over neutrino energy) providing insights of the  3--flavour structure. In Fig.~\ref{fig:kamland} the (later) beautiful result by 
KamLAND is reported, with almost two complete oscillation cycles observed.

To better explain the general picture it is necessary to go back to the initial idea of Pontecorvo, who in 1957 introduced the concept of 
neutrino oscillation~\cite{ponte-osci}, further elaborated by Z. Maki, M. Nakagawa and S. Sakata in 1962~\cite{mns} and Pontecorvo himself
in 1968~\cite{ponte}. However, one had to wait until the measurement of KamLAND for a clear understanding of the mismatch between the diagonalization of the charged lepton mass matrix and that of the neutrino mass matrix, similarly to what happens in the quark sector with the CKM
matrix~\cite{CKM}. The mismatch is described by a unitary matrix, $U_{\rm PMNS}$ to honor the pioneering authors, that mixes the 3 flavour states $\nu_{\alpha}$, $\alpha=e,\mu,\tau$,
of the weak interactions with the 3 mass eigenstates $\nu_{i}$, $i=1,2,3$:

\[ \left( \begin{array}{c} \nu_e \\ \nu_{\mu} \\ \nu_{\tau} \\ \end{array} \right) = 
\left( \begin{array}{ccc} U_{e1} & U_{e2} & U_{e3} \\ U_{\mu1} & U_{\mu2} & U_{\mu3} \\ U_{\tau1} & U_{\tau2} & U_{\tau3} \\
\end{array} \right) \cdot
\left( \begin{array}{c} \nu_1 \\ \nu_2 \\ \nu_3 \\ \end{array} \right).
\]
$U_{\rm PMNS}$ depends upon six real parameters: three mixing angles, $\theta_{12}$,  $\theta_{23}$, $\theta_{13}$ that
correspond to the three Euler rotations in a 3--dimensional space, and three phases, $\delta$, $\alpha_1$, $\alpha_2$. A suitable parametrization is
\[ U_{\rm PMNS}=
\left( \begin{array}{ccc} c_{12}c_{13} & s_{12}c_{13} & s_{13} e^{-i\delta} \\ 
-s_{12}c_{23} - c_{12}s_{13}s_{23} e^{i\delta} & c_{12}c_{23} - s_{12}s_{13}s_{23} e^{i\delta} & c_{13}s_{23} \\
s_{12}s_{23} - c_{12}s_{13}c_{23} e^{i\delta} & -c_{12}s_{23} - s_{12}s_{13}c_{23} e^{i\delta} & c_{13}c_{23} \\
\end{array} \right) \cdot
\left( \begin{array}{ccc} 1 & 0 & 0 \\ 0 & e^{i\alpha_1/2} & 0 \\ 0 & 0 & e^{i(\alpha_2/2)} \\ 
\end{array} \right)
\]
where $c_{ij} \equiv \cos\theta_{ij}$ and $s_{ij} \equiv \sin\theta_{ij}$. 
The phases $\delta$ ($\equiv \delta_{CP}$) and $\alpha_1$, $\alpha_2$ are Dirac--type and Majorana--type $CP$ violating phases, respectively.
The above description holds only for 3--flavours neutrinos, even if it can be extended to a basis
with one or more neutrinos, either {\em steriles}  i.e. neutrinos that do not couple to the weak interactions or neutrinos that do not 
contribute to the invisible width of the $Z^0$ boson. In that case one usually assumes
that extra neutrinos mix with the standard neutrinos, i.e. that the mixing matrix is not degenerate.
Applying the time evolution to the mass eigenstates in vacuum
\[ |\nu_i(t) \rangle = \exp(-iE_i (t)) |\nu_i \rangle ,
\]
and using the unitarity of the mixing matrix, the vacuum transition amplitudes and probabilities are obtained:
\[ P_{\alpha\beta} = A^{\ast}_{\alpha\beta} A_{\alpha\beta} = \sum_{i,j=1}^3 U^{\ast}_{\alpha i}U_{\beta i}
U_{\alpha j}U^{\ast}_{\beta j} e^{-i(E_i-E_j)t} \]
If ultra--relativistic neutrinos are taken then $E=\sqrt{p^2+m_i^2}\simeq E+\frac{m_i^2}{2E}$, with $t\simeq L$, and the transition
probabilities are expressed in terms of {\em frequencies}, defined as $\Delta m^2_{ij}L/E$, where $\Delta m^2_{ij}\equiv m^2_i-m^2_j$, $E$ is the neutrino energy\footnote{The issue on which energy to consider for the individual neutrino state
is analyzed and solved in~\cite{giunti-E}.} and $L$ the traveled distance.
In case of three flavours only two of such frequencies are independent, e.g. $\Delta m^2_{21} = m^2_2-m^2_1$  and 
$\Delta m^2_{32} = m^2_3-m^2_2$, usually named for historical reasons the {\em solar} (or $\delta m^2$) and the {\em atmospheric} (or $\Delta m^2$)\footnote{Sometimes, and usually in the phenomenological fits,  $\Delta m^2$ is defined as 
$\Delta m^2=m^2_3-\frac{m^2_2+m^2_1}{2}$.} oscillation frequencies, respectively. 
Finally, the evolution in time brings to probabilities for survival ($=$ 1 -- disappearance probability) and appearance  
of a neutrino flavour with energy $E$ over the distance $L$.
In the simplest case where only two--flavours are involved the probabilities are described by an oscillation {\em amplitude} that depends on the mixing angle
and an oscillation {\em frequency} that depends on the mass scales as well as the experimental constraints, $L$ and $E$. 
The two--flavour approximation was widely used
until 2002 when it was understood that this was not always appropriate.

Coming back to the CHOOZ/KamLAND results, the associated probabilities are obtained by convolution over the three flavours. 
If $\Delta m^2$ is expressed in eV$^2$, $L$ in km and $E$ in GeV, one gets (for a more exhaustive discussion
see e.g.~\cite{strumia}, chapter 3):
\[ \pnuenumu = s^2_{23}\sin^2 2\theta_{13}\sin^2(1.27\times \Delta m^2 L/E)+c^2_{23}
\sin^2 2\theta_{12}\sin^2(1.27\times \delta m^2 L/E),\]
\[ \pnuenue = 1 - \sin^2 2\theta_{13}\sin^2(1.27\times \Delta m^2 L/E) - c^4_{13}
\sin^2 2\theta_{12}\sin^2(1.27\times \delta m^2 L/E).\]
In the above expressions the CP--violating terms are omitted. Furthermore $|\Delta m^2|\gg \delta m^2$ is assumed, that is
$\Delta m^2_{31}\simeq \Delta m^2_{32}$, so removing for a while the issue on the neutrino--mass ordering\footnote{Mass ordering 
is associated to either $m_1<m_2<m_3$ or $m_3<m_1<m_2$. As far as oscillations
are concerned the dependences on the mass ordering come from the interference between two effects. In vacuum the interference can be
given by the joint atmospheric and solar oscillations (see later).}. 
The assumption
is  justified by the fact that atmospheric mass--splitting $|\Delta m^2_{32}|$ 
is more than one order of magnitude greater than the solar mass--splitting 
$\delta m^2_{21}$\footnote{Already towards the end of 2002, just after the SNO result, the two-flavour fits (see e.g.~\cite{fogli2002}) predicted the so--called large--mixing angle (LMA) as best fit with
a {\em solar} mass-splitting around $10^{-5}$ eV$^2$, much smaller than the interval indicated for the {\em atmospheric} mass--scale by the first measurement from 
SK, $5\cdot 10^{-4} < \Delta m^2 < 6\cdot 10^{-3}$ eV$^2$ at 90\% C.L., updated in 2004 as
$1.5\cdot 10^{-3} < \Delta m^2 < 3.4\cdot 10^{-3}$ eV$^2$ at 90\% C.L.~\cite{superk-2004}. The LMA was confirmed by including KamLAND results~\cite{fogli2003}.}. 

Analyzing the probability expectations for
\nuenumu\, at the nuclear--reactor energies, the results of CHOOZ and KamLAND are explained by a) the low value of $\theta_{13}$ that allows the first term to be dropped, 
and b) the low value of $\delta m^2$ and hence the necessity for a rather large distance $L$ to become sensitive to the 
second term (two orders of magnitude larger than the CHOOZ baseline of 1 km). 
Indeed Nature was very {\em vicious} with CHOOZ, since the value for $\theta_{13}$
was really ``around the corner'', and big enough to moderate the dreams/needs for new technologies. 
It is worth to outline that
CHOOZ could probably catch $\theta_{13}$ by lowering its systematics effects. 
However, researchers did not find sufficient
motivations to try to improve the detector just to gain few percents of phase--space. The next generation of reactor--experiments were developed under a general feeling of criticisms and disbelief. 

The experimental and phenomenological scenario after 2002 was really exciting: $\theta_{23}$ and $\theta_{12}$ had been measured whereas
the quest for the last missing mixing--angle $\theta_{13}$ prompted to a large variety of proposals, models and endless discussions. 
The possibility that $\theta_{13}=0$, which implies no
CP--violation in the leptonic sector, was widely parsed. That hypothesis had to be absolutely checked, even if very large--scale
detectors and accelerators were plainly needed. In that period flourish of ideas on new accelerating beams
like super--beams, beta--beams and neutrino--factories were developed (see e.g.~\cite{neu-fact}). Different strategies were set up, depending on the value of $\theta_{13}$,
below the limit set by CHOOZ, $\theta_{13}< 12^{0}$ (90\% C.L.). 
In retrospect, besides the 
novelty of new  techniques, always useful for future experiments, the lack of physics case (i.e. a very small $\theta_{13}$) seems nowadays evident. 
That is a lesson that the neutrino community should learn: the request for always larger detectors and systems should be really
motivated by a founded physics case. The prejudice that $\theta_{13}$ were very small was perhaps founded on the not proper appreciation of the
(excellent) data analysis performed by CHOOZ, believed more conservative than it really was. Many papers investigated
only the range to few degrees of $\theta_{13}$ (see e.g.~\cite{cao-old}). 
This point is further discussed in later Sections.

In the following 10 years, up to 2012, a large collection of measurements was gathered, all confirming the oscillation pattern
and (almost) all consistent with the 3--neutrino framework. For example, the very recent (2015) observation of the \nutau appearance
from a \numu beam by the OPERA experiment~\cite{opera-5}, was largely expected and, in some respects, just 
delayed of about 10 years\footnote{Proposal for the OPERA experiment dated 1997 and the project was finally
approved in 1999-2000, after the SK discovery on atmospheric neutrinos. Always in retrospect, it would have been wiser to stop the project
for a couple of years, and hence to cancel it. OPERA
may however be able to provide interesting insights on the presence
of sterile neutrinos at Long--Baseline (LBL)~\cite{palazzo-lbl,stanco-lbl}.}.

Besides $\theta_{13}$ and the possible leptonic CP--violation there were other missing parts, for which  the famous MSW effect
must be recollected.
In 1978 L. Wolfenstein~\cite{wolfe} showed that the propagation of neutrinos is significantly modified in the presence 
of ordinary matter due to their interactions with electrons, protons, and neutrons. Moreover, the coherent forward elastic scattering
amplitudes are not the same for all neutrino flavours, \nue, \numu and \nutau, since \nue have additional contributions due to their
charged current (CC) interactions with matter. After few years of studies (and corrections of mistakes)
by several authors, finally in 1985 S.P. Mikheev and A.Yu. Smirnov~\cite{sw} discovered resonant flavour transitions are possible
when neutrinos propagate in a medium with varying density. That briefly accounts for the electron--neutrino oscillation
pattern reported by the experiments on solar neutrinos. Similar effects are expected when neutrinos, in particular \nue, travel inside
the Earth at baselines of the order of at least 1000 km. 

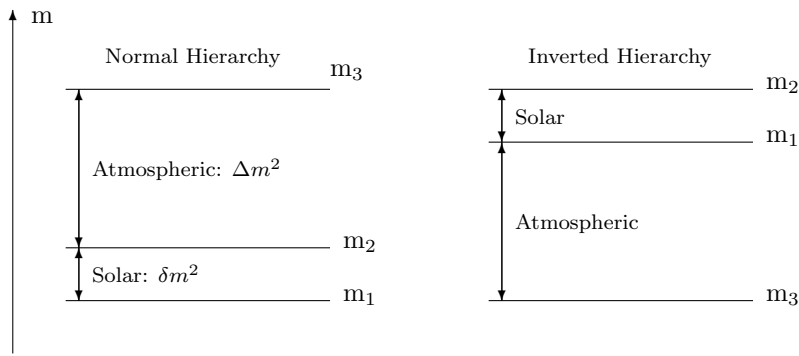
\begin{figure}[h]
\begin{center}
\begin{picture}(330,150)
\put(0,0){\vector(0,1){130}}\put(7,125){m}
\put(20,20){\line(1,0){100}}\put(20,40){\line(1,0){100}}\put(20,100){\line(1,0){100}}
\put(125,20){m$_1$}\put(125,40){m$_2$}\put(120,105){m$_3$}
\put(25,20){\vector(0,1){20}}\put(25,40){\vector(0,-1){20}}\put(25,40){\vector(0,1){60}}\put(25,100){\vector(0,-1){60}}
\put(35,110){\footnotesize Normal Hierarchy}\put(30,27){\footnotesize Solar: $\delta m^2$}\put(30,67){\footnotesize Atmospheric: $\Delta m^2$}
\put(180,20){\line(1,0){100}}\put(180,80){\line(1,0){100}}\put(180,100){\line(1,0){100}}
\put(285,20){m$_3$}\put(285,80){m$_1$}\put(285,100){m$_2$}
\put(185,20){\vector(0,1){60}}\put(185,80){\vector(0,-1){60}}\put(185,80){\vector(0,1){20}}\put(185,100){\vector(0,-1){20}}
\put(195,110){\footnotesize Inverted Hierarchy}\put(190,47){\footnotesize Atmospheric}\put(190,87){\footnotesize Solar}
\end{picture}
    \caption{Neutrino mass eigenstates for normal and inverted mass ordering (not to scale).}
    \label{fig:mh}
\end{center}
\end{figure}

An aside effect of solar--matter effects is that $\delta m^2 >0$. Instead, the sign of $\Delta m^2$
is not supplied, so far, by the atmospheric experiments. Thus, the structure of the neutrino mass--matrix is not fully determined,
and several solutions are possible when estimating the value of $\delta_{CP}$.
In the framework of three neutrino--flavours, the two possible solutions for the mass ordering are
usually named as Normal Hierarchy (NH) and Inverted Hierarchy (IH),  for $\Delta m^2>0$ and  $\Delta m^2<0$, 
respectively\footnote{Most often the definition for the mass hierarchy is defined as the sign of $\Delta m^2_{31}$ or
$\frac{1}{2}(\Delta m^2_{31}+\delta m^2_{21})$, which better reflect
the dependences in the transition probabilities. We keep $\Delta m^2_{32}$ for an easier interpretation in the present context.}
(Fig.~\ref{fig:mh}).

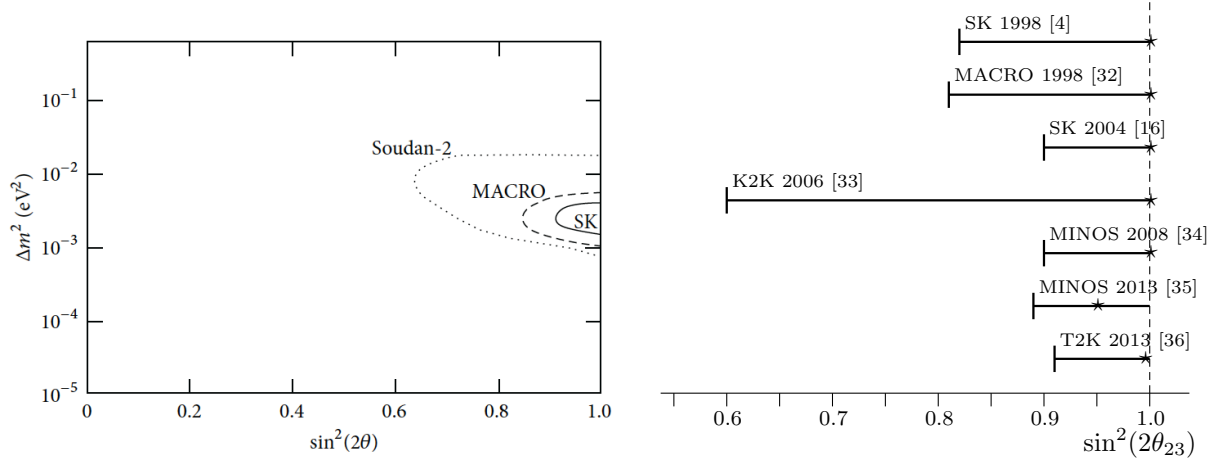
\begin{figure}[htbp]
  \includegraphics[width=0.5\textwidth]{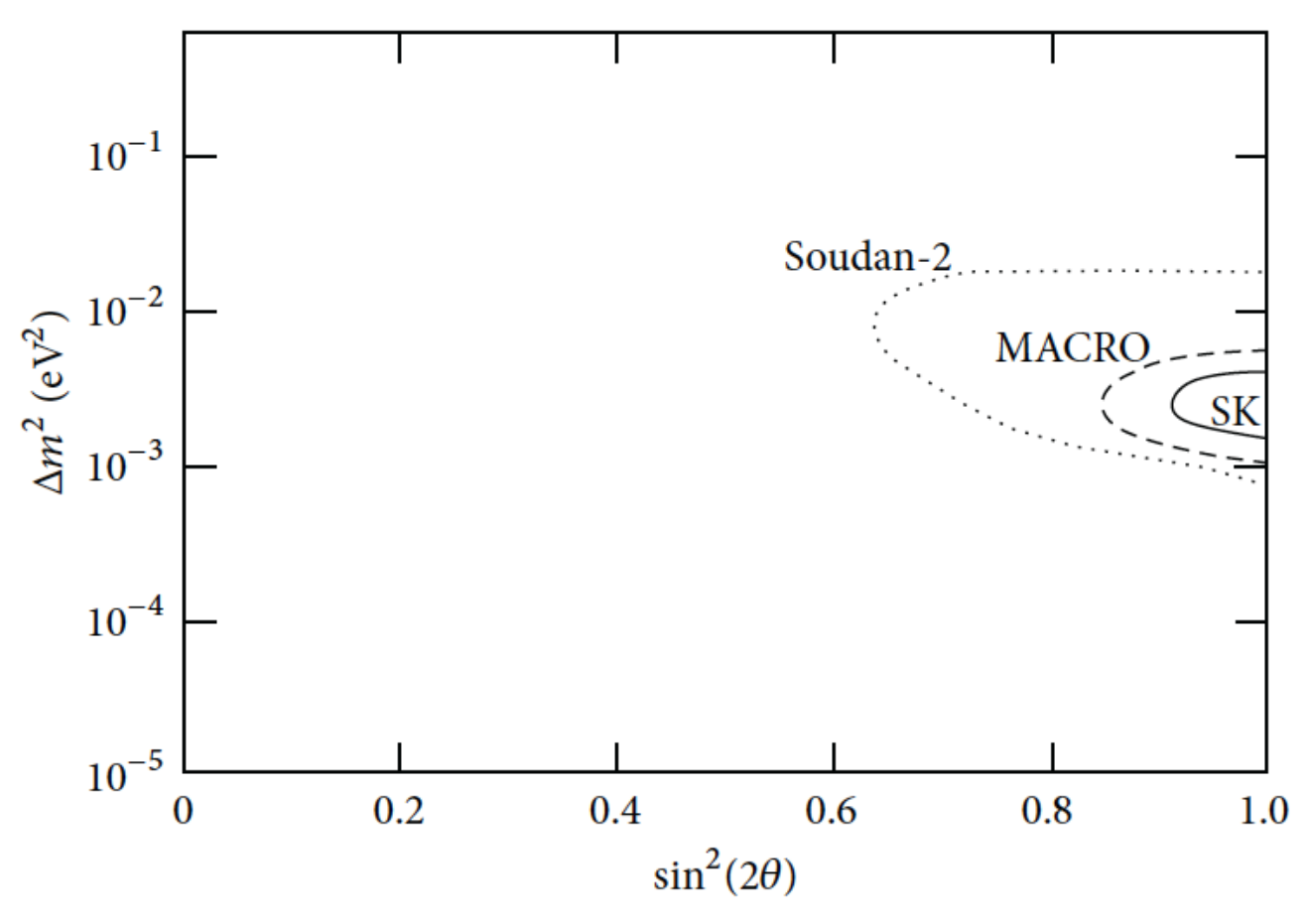}%
\begin{picture}(330,150)
\put(15,27){\line(1,0){200}} \put(20,22){\line(0,1){5}} \put(40,22){\line(0,1){5}} \put(60,22){\line(0,1){5}} \put(80,22){\line(0,1){5}} 
\put(100,22){\line(0,1){5}} \put(120,22){\line(0,1){5}}  \put(140,22){\line(0,1){5}} \put(160,22){\line(0,1){5}} \put(180,22){\line(0,1){5}}
\put(200,22){\line(0,1){5}}
\put(175,5){$\sin^2(2\theta_{23})$} \put(195,15){\footnotesize 1.0} \put(155,15){\footnotesize 0.9} \put(115,15){\footnotesize 0.8}
\put(75,15){\footnotesize 0.7} \put(35,15){\footnotesize 0.6} \multiput(200,27)(0,5){30}{\line(0,1){3}}
\linethickness{0.3mm}
\put(128,160){\line(1,0){72}}\put(128,155){\line(0,1){10}} \put(130,165){\scriptsize SK 1998 \cite{superk}}
\put(198,158){$\star$}
\put(124,140){\line(1,0){76}}\put(124,135){\line(0,1){10}} \put(126,145){\scriptsize MACRO 1998 \cite{macro}}
\put(198,138){$\star$}
\put(160,120){\line(1,0){40}}\put(160,115){\line(0,1){10}} \put(162,125){\scriptsize SK 2004 \cite{superk-2004}}
\put(198,118){$\star$}
\put(40,100){\line(1,0){160}}\put(40,95){\line(0,1){10}} \put(42,105){\scriptsize K2K 2006 \cite{k2k-2006}}
\put(198,98){$\star$}
\put(160,80){\line(1,0){40}}\put(160,75){\line(0,1){10}} \put(162,85){\scriptsize MINOS 2008 \cite{minos-2008}}
\put(198,78){$\star$}
\put(156,60){\line(1,0){44}}\put(156,55){\line(0,1){10}} \put(158,65){\scriptsize MINOS 2013 \cite{minos-2013}}
\put(178,58){$\star$}
\put(164,40){\line(1,0){36}}\put(164,35){\line(0,1){10}} \put(166,45){\scriptsize T2K 2013 \cite{t2k-2013}}
\put(196,38){$\star$}
\end{picture}
    \caption{Left: allowed regions for atmospheric neutrino oscillation,
    as measured by SK, MACRO and SOUDAN-2 at 90\%~C.L., from 1998 up to year 2004. The figure is taken from~\cite{gg}. 
    Right: the most relevant measurements
     of $\sin^2 (2 \theta_{23})$, at 90\% C.L., up to year 2013. The best fit, constrained to the physical region, is shown by the star.
     From around 2013 the angle $\theta_{23}$ began to be estimated with the $\sin^2 (\theta_{23})$ variable.}
    \label{fig:theta23}
\end{figure}

Together with $\delta_{CP}$ and the mass ordering, the third relevant ingredient for the evaluation of the full neutrino picture 
is the deviation, with sign, of the atmospheric mixing angle,~$\theta_{23}$, from~$\pi/4$. The mixing angles, 
$\theta_{ij}$ span two octants since $\theta_{ij}\in [0,\pi/2]$. While $\theta_{12}$ is around $33.5\degree$ within about 5\%, $\theta_{23}$
has always been measured compatible with maximal mixing, i.e. $\theta_{23}=\pi/4$. Currently its error is around 10\%.
The maximal mixing corresponds to an equal contribution of \numu and \nutau to the third neutrino mass state. Besides   
the solution of the mass ordering there are huge implications in cosmology and symmetry models: depending of their values
$\theta_{12}$ and $\theta_{23}$ regulate the percentage of  \nue, \numu and \nutau and the relative mass contributions.
It is interesting to consider at the historical series of measurements of  $\theta_{23}$, as reported in Fig.~\ref{fig:theta23}. 
From 2006 experiments on neutrino beams produced by accelerators, namely K2K~\cite{k2k-2006}, 
MINOS~\cite{minos-2008} and T2K~\cite{t2k-2013}, started to release results.
From around 2013 the improvements in the precision of the measurements forced the
analyses to be done within the full 3--flavours formalism, being also sensitive to its correlation with
MH (therefore $\theta_{23}$ started to be estimated with the $\sin^2 (\theta_{23})$ variable, see Fig.~\ref{fig:theta23}).
Presently, the best estimation is provided by T2K~\cite{T2K-2015-theta23}, followed by MINOS~\cite{minos-2014}.

Extended discussions on the neutrino oscillation in vacuum and in matter can be found nowadays in textbooks
like e.g.~\cite{giunti-book}, chapters 6 to 9. Some interesting papers on the historical perspectives, neutrino mass matrix 
and related issues are~\cite{gg}, \cite{king} and~\cite{verma}
(we do purposely avoid to list notable papers
before the measurement of 
$\theta_{13}$ in 2012 because the outlook differs before and after that date).

After the years 1998 and 2002 the third {\em annus mirabilis} for neutrino physics was 2012. Predictions of a large value
for $\theta_{13}$, i.e. close to the CHOOZ limit, were made in 2011 as preferred solution of the phenomenological fits~\cite{fogli2011},
just before the discovery was claimed by the reactor experiments in 2012~\cite{dayabay}.
The current estimated value of $\theta_{13}$ is $8.5\degree $~\cite{theta13} with a combined 5\% precision at 1 $\sigma$ level.

With the assessment of a non--zero and relatively large value of $\theta_{13}$ the possibility to measure CP--violation in 
the leptonic sector in a reasonable period of time is highly increased. In fact, the CP violation arises from the complex phase of the mixing matrix ($\delta_{CP}\ne 0,\pi$) and from the presence of at least three flavours that mix up ($\theta_{13}\ne 0$). The measurement of $\delta_{CP}$ may come from the different transition probabilities for neutrinos and antineutrinos\footnote{While
the $CP$ conserving terms depend on $\sin^2(1.27\times \Delta m^2 L/E)$, and it is the same for neutrinos and antineutrinos, 
the $CP$ violating part depends on $\sin(2\times 1.27 \Delta m^2 L/E)$. The latter oscillates with a doubled frequency compared to the $CP$ conserving part, being of opposite sign for neutrinos and antineutrinos. The need for
at least three flavours is simply due to the fact that, even if one introduces an additional $CP$ phase, the quartic invariant of 
the transition amplitude cannot become
complex for two flavours. That is the reason, in passing, that forced Kobayashi and Maskawa to postulate three flavours in the quark sector due to the presence of $CP$ violation in the hadronic flavour mixing.}. However, if the $CPT$ invariance holds, the transition probabilities
for {\ensuremath{\nu_i \rightarrow \nu_j} and for {\ensuremath{\overline{\nu}_j \rightarrow \overline{\nu}_i} are equal. Hence
no $CP$--violation can be observed in the disappearance mode ($i=j$). Instead, one needs to observe the transition among flavours, for both neutrinos and antineutrinos, to access CP violation.

The disentangling of the full picture in neutrino flavours/oscillations might come from detailed global fits, which are able to
cover the whole set of available measurements. In such a case only joint analyses that include the different outcomes from all the 
experiments would produce reliable results. This sentence may appear too much blunt, in particular to experimentalists, who 
vigorously analyze their data to extract physical results. Our conclusion is motivated by the fact that either the neutrino interactions
collected by single experiments usually provide quite limited datasets or
correlation terms of the oscillations become too much large to be neglected.
One example of analyses developed under questionable approximations, using only data of its own experiment, can be found
in~\cite{icarus-sterile,opera-sterile}. The consistent analysis in that case is the one performed in~\cite{palazzo-lbl}.
The correct approach seems finally accepted by the community, see 
e.g.~\cite{lisi-2015,valle-2014,schwetz-deltacp,palazzo-2015}. 
However, for the author, there are still some points that need to be better clarified. Between them the most relevant one is the way to 
establish the mass hierarchy. It will be treated in the next Section.

We end this short review on neutrino phenomenology by addressing the fourth point in the list of critical parameters to apprehend
in the near future, the sterile neutrino. The experimental
story of sterile neutrinos began in 1998 with the results of the LSND experiment~\cite{lsnd}\footnote{Actually, the first publication of LSND
dated 1995~\cite{lsnd-1995} and it immediately addressed the $\Delta m^2$ region around 1~eV$^2$. The possibility
of a sterile origin was part of the discussion afterwards. However, the {\em sterile}
hypothesis was seriously considered only after the discovery of the atmospheric oscillation in 1998, when 
it became clear that the LSND result did not really fit with both that observation and the solar neutrino deficit~\cite{giunti-1999}.}. 
At that time these results generated some confusion on global analyses, studies and proposals. 
Unfortunately the only experiment setup to confirm the result of LSND was inconclusive~\cite{mini-mu,mini-sci-mu}.
At the same time the quest and request of (at least one) sterile neutrino rose up, in particular for its possible contribution
to dark matter. Furthermore, from around year 2010, additional experimental hints emerged from computations of
the reactor neutrino fluxes~\cite{reactor-sterile}
and the calibration of radioactive sources~\cite{source}. These neutrino {\em anomalies} could be coherently 
interpreted as due to the existence of a fourth sterile neutrino with a mass at the eV scale.
Thus there were/are sufficient motivations to develop more projects and proposals~\cite{whitep}. 

So far, there is no demonstration of a sterile neutrino state with a mass around 1 eV.
The current projects are attacking the issue mainly from three sides: oscillation behavior at short distance (SBL beams)~\cite{fermi-sbl},
deficit of \nubare at nuclear reactors~\cite{reactor-new}, and \nue and \nubare disappearance from radioactive sources~\cite{source-new}. Even if the real possibilities of these proposals remain challenging, they would
either confirm or disprove the LSND result. Another proposal~\cite{nessie-prop} 
being at the same time robust and able, in case of a positive outcome,  to
fully demonstrate the {\em sterile} origin of the anomalies (as originally addressed in~\cite{stanco-muapp}) was unfortunately not
approved by FNAL. 

The presence of sterile neutrinos, in particular at 1 eV, is a very open question that affects the results of the
analyses. A good example is in~\cite{palazzo-2015}, where the inclusion of a sterile state is shown to wash out the disentangling 
of the mass hierarchy.
The final (negative) response on sterile neutrinos of 1 eV mass is expected in the next couple of years (by the year 2017--2018), when reactor
and radioactive source experiments will start their (short) data taking.
It is worth to outline that 
measurements sensitive to the possible presence of sterile neutrinos are also
expected by IceCube~\cite{icecube}
and LBL experiments,  like e.g. MINOS$+$~\cite{minosp}.

It is clear that many relevant questions on neutrinos have not been considered here, since we focussed mainly on 
neutrino flavour oscillations. However it is important at least to mention the neutrinos from supernova bursts, the direct
mass measurement, the solar and cosmological contributions of neutrinos. These four areas of investigations are undoubtedly relevant
i.e. they are all worth {\em per se} the current and planned future activities. Other projects, not mentioned here, are probably not worth major effort, being interesting only as side results of more general items.

\section{Discussion on the MH determination}\label{sec:mh}

The issue of the mass ordering has been highly debated in the last decade, but it gained in interest
with the discovery of the relatively large value of $\theta_{13}$ in 2012. The convolutions between the three mixing angles and 
the mass parameters are such that  measurements of the current experiments may become sensitive to the 
dependences of the oscillation probabilities to the sign of MH. 
Surely the MH determination will be a major point for the next experiments under construction.
All the methods developed for establishing whether MH is normal or inverted are based on the computation of the difference
of $\chi^2$ with respect to the best--fit solutions of NH and IH (\cite{mh-qian,mh-ciufoli,mh-schwetz}). 
Even the Bayesian--statistics approach finally deals with that {\em indicator}~\cite{mh-blennow-bayes}. The  adopted expression is:
\[ \Delta\chi^2= {\rm min}\chi^2_{IH}-{\rm min}\chi^2_{NH}, \]
where the $\chi^2$, computed for a set of parameters, is defined from the likelihood $\cal{L}$ of the data according to $\chi^2=-2\log(\cal{L})$,
thus making $\Delta\chi^2$ equivalent to the likelihood ratio of the best fit points in IH and NH cases.

In the frequentist approach several unappealing features are present, like in some cases the rejection of both NH and
IH, and the fact that the chosen statistics cannot be approximated by a Gaussian one, or, in other words, Wilks' theorem~\cite{wilks} does not apply to a binary measurement. In the referred papers the second issue has been solved,
the first has not. Moreover,
the way the sensitivity is computed (how many $\sigma$'s and which test--statistics is used) is matter of 
discussion. 
Further concerns relate to a more general consideration.
So far in all the computations the interplay
between the two options, NH and IH, has been taken into account by computing the best--fit solutions 
for both options\footnote{I saw only one study done by SK~\cite{mh-sk} that assumed as true the NH and computed the $\chi^2$ for IH based on
the best solution of NH. This goes along the approach suggested here, but it is incorrect as it does not exclude the
possibility of degeneracy in case IH owns a different solution with a $\chi^2$ at a similar level.}. In this way they provide an answer
to the question: what is the right hierarchy? The procedure mimics what has been developed for the Higgs search~\cite{cowan,stat-higgs,higgs},
based on the so--called $CL_s$~\cite{read} method, and that has been also positively applied in the search for sterile neutrinos~\cite{nessie,daya-bay-sterile}. However, if the neutrino phenomenology
is described within the 3--flavour pattern, one of the two options has to be true and the other wrong, i.e. NH and IH are 
mutually exclusive. From a physical point of view we are not really interested in the wrong answer.
In other words, we should prefer to identify the right hierarchy, forgetting about the wrong one. It is a discovery process, and not an exclusion one.

The confusion arose from the fuzzy evidence on the sign of MH that began
to appear in 2012. It was believed more efficient/right to use statistical estimators sensitive to both options: [NH true -- IH false] and 
[NH false -- IH true], as exhaustively discussed in~\cite{mh-qian}.
This approach may be justified for analyses based on set of data coming from a single experiments, but in this case the sensitivity is a quite delicate
issue, as discussed in~\cite{mh-schwetz}\footnote{Anyway, the approach on $\Delta\chi^2$ brought to uncorrected
conclusions grounded in the discussion of type I and type II errors, which is not appropriate for the physics case under discussion.
The neutrino community was led to think, perhaps by some of them trying to promote a (too much) robust experimental proposal, 
that the accurate sensitivity on MH should be extracted only via the contemporaneous evaluation of the two types of errors.
That is right only when the two hypothesis are alike expected (!) and their PDF's  overlap in the phase space.}. 
It is interesting to note that in the last three years confusion and disputes about the sensitivity on MH have been so large that 
some papers~\cite{mh-lisi-1} even quoted the sensitivity for just the right solution itself.
In the light of the present discussion that choice corresponds to the worst option. It would have been similar to establish 
the Higgs particle just by counting the initial handful of collected events, so reaching a 2--3 $\sigma$ sensitivity instead of 5 $\sigma$.
However the idea in~\cite{mh-lisi-1} 
to introduce a continuous variable $\alpha$ might be interesting since it would allow to perform an analysis \`a la 
Higgs~\cite{stat-higgs}. One should just keep in mind to evaluate the sensitivity to disprove IH, i.e. to compute
the minimization starting from $\alpha=-1$ following the notation of paper~\cite{mh-lisi-1}. 
It is worth noting that~\cite{mh-lisi-1,mh-japon} took the approach that is going to be proposed below.
Unfortunately the same authors decided to go for the $\Delta\chi^2$ estimator in their recent paper~\cite{mh-lisi-2}\footnote{If 
people prefer to stay with the comparison between IH and NH, a surely better solution than the usual difference
in $\chi^2_{\rm min}$ is the evaluation of a modified $F$--test, where the test statistic is defined as  
$F=({\rm min}\chi^2_{IH}-{\rm min}\chi^2_{NH})/{\rm min}\chi^2_{NH}$.  $F$ follows a Fisher distribution
when the two compared models cannot be 
statistically distinguished. From the $F$--test a $p$--value can be extracted and the sensitivity computed in terms of $\sigma$'s 
(in the one--sided procedure!). A further advice is to use always $-2 \log {\cal L_{\rm max}}$ instead of $\chi^2_{\rm min}$ 
since there are non--Gaussian distributions in the set of random variables (that was not always done in the literature).}.

We suggest that today the issue can be approached in a more basic and straightforward way, more comfortable and more understandable for the community. A change in perspective is therefore desirable.
In particular one must decide to investigate the sensitivity to either confirm or discard one of the two MH's. As experienced in many
past measurements, it is usually much simpler to evaluate the discovery of a signal than to quantify its exclusion
with an upper--limit. 
The key--point is to identify a good, and possibly optimal estimator for the test statistics.
In this context the discovery of a new signal is founded on the exclusion of the no--signal hypothesis ($H_0$), and its sensitivity is 
given by the $p$--value of the test statistics on $H_0$. There is no need to include information on the signal hypothesis ($H_1$).
The case of the Higgs search is somehow different. Looking for new particles the strength parameter $\mu$ that weights the
cross--section of the new particle is introduced. The $CL_s$ method is an optimal one~\cite{read}, and $\mu$ is tested against the $\mu=0$ hypothesis
over the best solution $\hat{\mu}$ when looking for discovery. For NH/IH case an optimal test statistics has to be looked for. That cannot
be $\mu$, simply because the best fit is obtained over the sum {\em signal} $+$ {\em background}, while NH and IH are 
mutually exclusive hypotheses.

It is also mandatory to check the consistency of the data among themselves. For example, if the estimation 
of $\theta_{13}$ by reactor experiments would not be consistent to that by LBL ones, let us say for less than  10\% on the related
phase space, it should make no sense to put them together to extract a solution for MH. 

In the very recent release of results from NOvA~\cite{nova1,nova2} the right approach was chosen: the significance on MH has been
computed separately for NH and IH, taking $\theta_{13}$ as estimated by reactor data. Despite
the not so precise wording and the use of the biassed technique of 
Feldman\&Cousins~\cite{Fel-Cou}\footnote{It is well know that the Feldman\&Cousins technique is biassed 
by the amount of level of background~\cite{read}.}, we can extrapolate from 
Fig.~\ref{fig:nova-mh} (from figure 4 of~\cite{nova2}) that NOvA disfavors IH at 3 $\sigma$  (double--sided?) in the 
$0<\delta_{CP} <0.8\, \pi$ range, for the less restrictive data selection. 

\begin{figure}[htbp]
\begin{center}
  \includegraphics[width=0.6\textwidth]{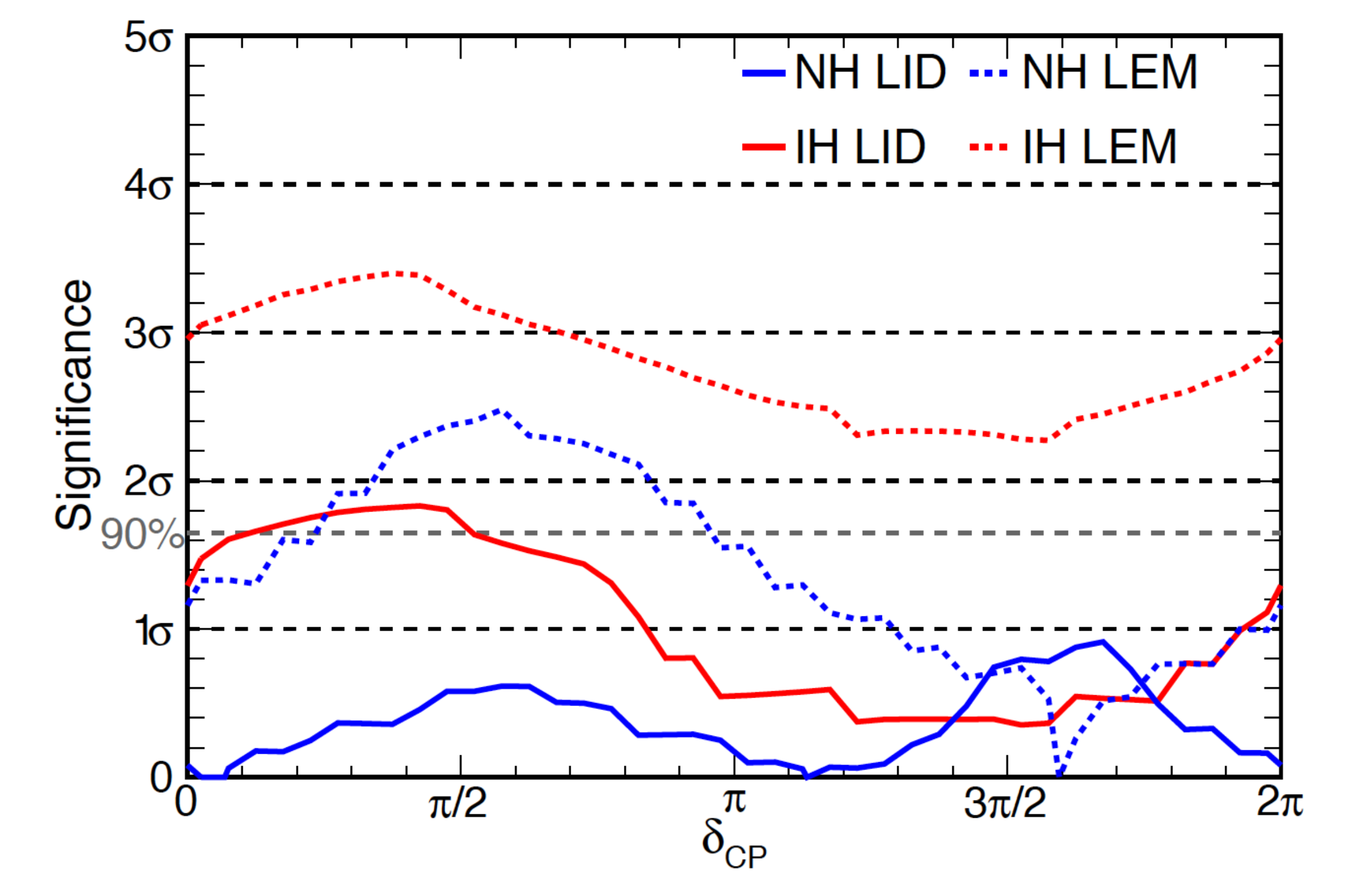}%
    \caption{The important result from NOvA on the statistical significance of MH as function of $\delta_{CP}$ 
    (figure 4 of~\cite{nova2}). $\theta_{13}$ is constrained by the reactor measurements. Blue curves are obtained when the 
    more restrictive data
     selection is applied, while the red curves  corresponds to the less restrictive data selection. This result, if correctly interpreted, may become the turning point for the future.}
    \label{fig:nova-mh}
\end{center}
\end{figure}

Since there are already hints and intuitive arguments that the option chosen by Nature be NH, the right approach should 
establish the exclusion of the IH possibility. Then, a simple goodness--of--fit test would be sufficient to disprove IH.
Groups involved in global fits should start to quote the ${\rm min}\chi^2_{IH}$ and its probability,  quantifying the $p$--value
for IH, properly defined as $P(\chi^2\ge{\rm min}\chi^2_{IH})$. It is worth to note that the conversion to the number of $\sigma$'s 
has to be computed via the one-sided Gaussian--test convention, since the minimum $\chi^2$ is computed\footnote{The distinction between one--sided and double--sided becomes not relevant when the 3 $\sigma$'s limit and above is reached.}.
In~\cite{mh-stanco} a new estimator for MH is introduced as well as the subsequent perspectives for the determination of the mass hierarchy in the near future.

Following the NOvA results an interesting work was developed by A. Palazzo (\cite{palazzo-2015}), who took into account 
also the effect of a sterile neutrino at the 1 eV mass scale. In the 3--flavour picture a consolidation of the preference for NH is extracted,
disfavoring CP conservation with a statistical significance close to 90\% C.L.. However the preference is washed out
in the 3$+$1 framework. The light sterile neutrinos constitute a potential source of fragility in the identification of the 
right mass hierarchy, demonstrating the need of clarifying the sterile issue as soon as possible.
We will address some further issues on the determination of MH in the conclusive Section of the paper.

\section{Relevant neutrino experiments for the near future}

We will describe shortly the important features of two currently running experiments, T2K and NOvA, together
with the already under construction JUNO experiment.

\vskip10pt
$\underline{T2K\, experiment}$
\vskip 5pt
T2K, a long--baseline experiment proposed in 2003~\cite{t2k-prop}, had initially three clear goals, namely to measure $\theta_{13}$,
to refine measurement of $\theta_{23}$ and to search for sterile components in the \numu disappearance mode.
Starting the data taking in 2009, T2K beautifully accomplished the first two tasks few years 
later~\cite{T2K-theta13-2014,T2K-2015-theta23}, while it was not
appropriate for the sterile search lacking the second detector at close distance. Indeed T2K includes SK as large detector
at 295 km far and only one very close detector to the neutrino beam of the J--PARC facility. Its results (e.g.~\cite{T2K-2015-theta23}) were obtained with
a power steadily increased up to 220 kW that allowed the collection of $3.01\times 10^{20}$ protons 
on target (POT). The \numu beam, peaked at 0.6 GeV with a $\sim$ 0.2 GeV wide--band is 4.4 mrad off--axis the SK detector. 
Systematic errors due to the detection (different targets at near and far sites) are currently the largest ones. Proposals to lower
the systematics at 2--3\% level are under scrutiny.

Even if conclusive results on the sterile neutrinos are missing, T2K is now in the optimal position for a substantial
contribution to the measurement of the mass hierarchy and the CP phase~\cite{T2K-ptep}. That potentiality has been greatly
recognized, and on July 15th the J--PARC Physics Advisory Committee approved the upgrade of the Main--Ring and a new plan aiming to reach 900 kW in 2020 (from the current 370 kW). The overall foreseen exposure ($7.80\times 10^{21}$ POT), to be collected in a time--schedule of 5 years, is more than 20 times larger than what used until now for the published results.

Last summer the first study of \nubarmunubare was 
reported by T2K~\cite{T2K-antinue}. Despite the handful of events not yet sufficient to discriminate background, it is 
highly expected that, depending on the value of  $\delta_{CP}$ phase, a stronger evidence will be reached in one or two more years of data taking, starting to probe in a {\em direct} way\footnote{The result on \numunue at LBL depends tightly on both
$\delta_{CP}$ and $\theta_{13}$. By combining it with the $\theta_{13}$ measured at the reactors, which are practically independent
of $\delta_{CP}$, indirect information on $\delta_{CP}$ are extracted. By combining \numunue and \nubarmunubare at LBL
a direct measurement can be obtained.}  the CP symmetry in a LBL experiment.

\vskip10pt
$\underline{NOvA\, experiment}$
\vskip 5pt
NOvA~\cite{nova-exp} is a multi--purpose experiment located 14 mrad off--axis at 810 km from the neutrino beam source, NuMI (FNAL).
The huge far detector has a mass of 14 kton, with 8.7 kton of active liquid scintillator mixed with oil.
A similar detector, 0.3 kton, quite close to the beam, is also part of the system. Thanks to the off--axis position the two NOvA detectors
can collect neutrino interactions at 2 GeV energy with a spread of about 0.25 GeV.
Very efficient measurements of \numu and \nue interactions are available. In particular \nue CC candidates are 
selected in the restricted (released)  $1.5 (1.3) < E< 2.7$ GeV window, with expected background
as low as 10\%. The systematic effects related to the (very similar near and far) detectors are not relevant, while the major limitations come
from the current knowledge of neutrino parameters, including its cross--sections.

NOvA was commissioned in 2014 and at the same time it started to take data. First results, based on $2.74\times 10^{20}$ POT were released in 
August this year~\cite{nova1}, 
providing solid evidence of \numunue oscillations.
With \nue and \nubare data NOvA can obtain results on the whole set of interesting parameters, namely MH, 
$\delta_{CP}$ and the octant of $\theta_{23}$. The overall foreseen exposure is 13 times larger, to be collected in about
6 years time--schedule, half $\nu$ and half \nubar.
The NuMI beam reached already the power record of 520 kW. Major upgrades are foreseen that will bring the power operations
at 700 kW.

\vskip10pt
$\underline{JUNO\, experiment}$
\vskip 5pt

The JUNO experiment consists of 20 kton target mass of Linear Alkyl-Benzene liquid scintillator, monitored by about 
18000 twenty--inch high--QE photomultipliers providing a ~80\% photo--coverage. It will allow an unprecedented 3\%$/\sqrt{E}$ 
energy resolution at 1 MeV to detect \nubare coming from reactor plants of about 20 GW power, at a first stage, 53 km far away.
Approved in 2014, it is foreseen to start data taking in 2020. The current conceptual design report is in~\cite{juno-cdr} 
and the physics achievements are described in~\cite{juno-physics}. 

With regards to the the confirmation of MH at high significance, Juno will be relevant only if the time window 
2020--2025 is kept on schedule. JUNO will allow to single out one of the missing fundamental parameters, MH, in an almost independent (and different) way of the others (no dependence on $\delta_{CP}$, no strong dependence on three-- vs four--neutrino pattern, no dependence on $\theta_{13}$,
no dependence on matter effects). 
It is revealing that if the standard picture holds then the relevance of such confirmation on MH would not be so high
(remember the case of the OPERA \nutau). 

Indeed, more than results on MH and increasing the accuracy on $\theta_{13}$ from 7\% to 5\% or better, not so interesting
for the constraint on MH, JUNO will substantially reduce the uncertainties on the solar oscillation parameters, namely on $\theta_{12}$ and $\delta m_{12}$,
by almost an order of magnitude in six years of data taking (see Section 3 of~\cite{juno-physics}).
Therefore, the global significance on MH, and $\delta_{CP}$, will significantly increase\footnote{A similar
improvement in the MH significance will be also given by the increase precision on $\theta_{23}$ that will be provided by T2K
and NOvA themselves.}. 
In any case JUNO, as the third big multi--purpose experiment, will join the team of T2K and NOvA, and they will beautifully crown the 
invaluable results to be gained within the next decade on neutrinos, i.e. for the year 2025.

\section{Summary}

Neutrino oscillation studies are entering the era of precise and refined measurements, mainly due to the 
success of the 3--flavour formalism. To confirm the scenario only three parameters have still to be determined with the
sufficient precision, namely, the mass hierarchy of the neutrino mass states, the value
of the CP phase and to which octant belongs $\theta_{23}$. If the present hints for a Normal Hierarchy were confirmed by the 
next datasets to be collected by the current long--baseline
experiments, T2K and NOvA, and if the already in construction JUNO experiment  promptly released results in due time, 
a final disentangling of the full picture would probably be achieved. We also need Nature be not too devious. 
Its {\em deviousness} if any should not received too much weight
in view of the enormous costs that new projects have to face. The history of the recent physics discoveries, in particular
in the field of neutrinos, is there to show that researchers should not be more weird than Nature.

There is however the possibility that this scheme be proved wrong/incomplete if a sterile neutrino state at 1 eV mass scale exists.
That has to be clarified as soon as possible. A negative answer can be obtained quite soon by reactor and
radioactive--source experiments. In case
the answer were inconclusive or even positive, quite new scenarios have to be considered and fully new proposals 
be developed. 
Finally, an exciting period of data collection and analyses is foreseen in the next three to five years (by the year 2018--2020),  provided the right approach be 
taken, as exemplified below for the mass hierarchy.

\vskip10pt
$\underline{Mass\, Hierarchy}$
\vskip 5pt
The determination of the mass hierarchy is likely to be obtained in the near future. In Section~\ref{sec:mh}
we briefly summarized the active studies performed mostly in the last three years. We concluded arguing that a new basic approach
should be taken for the determination of MH. Provided that the current hints for NH are confirmed (a) by the next data 
on \numunue by NOvA~\cite{nova1} and T2K~\cite{T2K-antinue}, (b) by the potential contribution of new data analyses from 
the atmospheric neutrinos
and (c) by the outcomes of the next global fits, and that all these new information are consistent among themselves
at a C.L. as large as 90\%, then the sensitivity on NH should be computed following the approach to disprove IH. 

What would be the sufficient level of
sensitivity is already matter of discussion. The usual rule of thumb in HEP is to consider a 5 $\sigma$ level, and many
proposals advised that value (e.g. the recent DUNE and HyperK studies). However this assumption is entirely {\em a priori}. Some recent papers already expressed some concerns about the reasonableness of that choice for the neutrino MH~\cite{mh-may2015}. 
The rule of 5 $\sigma$ for discovery of new particle states
is an excellent way to establish the existence of new particles. Reason for that stands in the long history of particle physics 
(from about 1980), which proved that is the right value to choose, at least when the statistical analysis is properly done. 
Many signals at 3--4 $\sigma$
were not  eventually confirmed because of the Look--Elsewhere--Effect (LEE) and/or some hidden systematics. There are 
very few and very  criticized examples where the systematics effects were so large, and so naively missed by researchers,
that even the 5 $\sigma$ rule was unacceptably disproven. In any case such examples occurred in individual 
experiments\footnote{The famous case of pentaquarks around 2003--2006 was actually different. Many experiments looked at the possible
presence of a new state around 1.5 GeV, with about half in favor and half against its existence. On top of an evident 
bandwagon effect, the statistical analysis was somewhat poor, at least in the first papers, and anyhow the controversy has not been
clarified yet~\cite{leps}. The recent discovery of pentaquarks by LHCb~\cite{lhcb} may re--open the issue for the
lower energy $s$--content pentaquarks.}.

The situation is quite different for MH. There is obviously no LEE, the statistical issue is already extensively discussed, and, most
relevant, results will come from the combinations of several different experiments and different physics channels. The only  
assumption will be the 3--flavour neutrino pattern. Hence, we tend to conclude that already a 3 $\sigma$ level, i.e. a probability
of 99.73\% in the bayesian interpretation, would be sufficient. In fact what is meant with ``sufficient level'' is the persuasion
that the result is right. For a new particle state a 5 $\sigma$ level is needed. For a binary option, grounded on the assumption that
the 3--flavour neutrino is the right model, a 3 $\sigma$ level should be sufficient. Otherwise, to advocate higher levels one should 
address  the possible origins of mistakes that would disprove the result.

To conclude this issue it can be ascertained that in three, maximum four years from now, pending the efficient running
of NoVA and T2K and the confirmation of the current hints, the MH should be resolved~\cite{mh-stanco}. The next generation of experiments, i.e. the JUNO experiment that is foreseen to start to
take data in 2020, would confirm the result mainly by achieving better resolutions on the other
parameters of the neutrino matrix (especially $\theta_{12}$ and $\delta m^2$) rather than by the direct measurement of MH.
Nevertheless, the possible measurement of MH by JUNO would address different parts of  
the oscillation pattern, in particular there will be no dependence on the matter effect, therefore strengthening the global
picture.

Unfortunately ICAL/INO is foreseen to reach a good sensitivity for MH in 10 years of running, i.e. in 2030. 
If the expected result on MH should be confirmed, ICAL would arrive too late 
for any quantitative contribution to this topics. 
On the same line the physics case for DUNE and HyperK, for what concerns MH, would be completely washed out in
three years from now. The related issue on $\delta_{CP}$ would be also heavily affected, and DUNE/HyperK might totally miss the 
physics--case for these neutrino studies. Nevertheless, any incoherence that could originate between results from different experiments, 
validated by tensions arising from the global fits, would confute such conclusions.
These arguments will be reported with more details and in a quantitative way in~\cite{mh-stanco}.

\vskip10pt
$\underline{Conclusion}$
\vskip 5pt
Neutrino Physics is a very attractive field of interest due to its role in the Standard Model and its still unknown parameters.
In 2012 the discovery/measurement of a relatively large $\theta_{13}$, the third neutrino mixing angle, had a two fold consequence: the first was to open
a new era of precise determination of parameters, coherently  described in the 3--flavour oscillation pattern; the second was just
to increase the belief in the oscillation paradigm itself (worth to remind that the value of $\theta_{13}$ was foreseen with good precision by global fits). 
The two large experiments currently taking data, T2K and NOvA, will be able to collect a big harvest of results.
The concurrence of JUNO that is foreseen to start taking data in 2020 would complement them.

The strategy for the next generation of experiments should necessarily take note of the results that will be reached in the next three to five years (by the year 2018--2020).
From this point of view western countries should learn from the past and try to avoid non--optimal choices for the next future, as happened to Europe  in the past two decades.

\section*{Acknowledgements}
\label{sec-ack}

I  thank S. Dusini, A. Longhin and L. Patrizii for, on top of the reading of the manuscript and their suggestions,
the enjoyable discussions we had about the issues discussed in this paper.


\begin{thebibliography}{9}
\bibitem{davis} B.T. Cleveland et al., Astrophys. J. 496, 505 (1998).
\bibitem{bahcall} J.N. Bahcall, A.M. Serenelli and S. Basu, Astrophys. J. 621, L85 (2005) [astro-ph/0412440].
\bibitem{bahcall-2} J.N. Bahcall and R. Davis, Science 191, 264 (1976).
\bibitem{superk} Y. Fukuda et al. (Super-Kamiokande collaboration), Phys. Rev. Lett., 81, 1562 (1998) [hep-ex/9807003].
\bibitem{sno} Q. R. Ahmad et al. (SNO collaboration), Phys. Rev. Lett., 89, 011301 (2002) [nucl-ex/0204008].
\bibitem{chooz} M. Apollonio et al. (CHOOZ collaboration) Phys. Lett., B420, 397 (1998) [hep-ex/9711002].
\bibitem{kamland} K. Eguchi et al. (KamLAND collaboration), Phys. Rev. Lett., 90, 021802 (2003) [hep-ex/0212021].
\bibitem{kam-fig} S. Abe et al. (KamLAND collaboration), Phys. Rev. Lett., 100, 221803 (2008) [arXiv:0801.4589v3].
\bibitem{ponte-osci} B. Pontecorvo, Sov. Phys. JETP, 6, 429 (1957).
\bibitem{mns}  Z.~Maki, M.~Nakagawa, and S.~Sakata, Prog. Theor. Phys. 28, 870 (1962).
\bibitem{ponte} B.~Pontecorvo, Sov. Phys. JETP 26, 984 (1968).
\bibitem{CKM} N. Cabibbo, Phys. Rev. Lett. 10, 531 (1963);\\
M.Kobayashi and T.Maskawa, Prog. Theor. Phys, 49, no. 2, 652 (1973).
\bibitem{giunti-E} C. Giunti and C.W. Kim, Phys. Rev. D58, 017301 (1998) [hep-ph/9711363].
\bibitem{strumia} A. Strumia and F. Vissani, arXiv:hep-ph/0606054v3 (2010).
\bibitem{fogli2002} G. L. Fogli et al., Phys. Rev. D66, 093008 (2002) [hep-ph/0208026].
\bibitem{superk-2004} Y.Ashie et al. (Super-Kamiokande collaboration), Phys. Rev. Lett. 93, 101801 (2004) [hep-ex/0404034];
Phys. Rev. D71, 112005 (2005) [hep-ex/0501064].
\bibitem{fogli2003} G. L. Fogli,  E. Lisi, A. Marrone and D. Montanino,  Phys. Rev. D67, 093006 (2003) [hep-ph/0303064].
\bibitem{neu-fact} C. Albright et al., {\em The Neutrino Factory and Beta Beam Experiments and Development}, 
arXiv:physics/0411123v2 (2004);
T. Abe et al., {\em Detectors and flux instrumentation for future neutrino facilities},  arXiv:0712.4129 (2007);
J. Bernabeu et al., {\em EURONU WP6 2009 yearly report: Update of the physics potential of Nufact, superbeams and betabeams}, arXiv:1005.3146 (2010);
R. J. Abrams et al., {\em Interim Design Report},  arXiv:1112.2853 (2011);
S. Bertolucci et al., {\em European Strategy for Accelerator-Based Neutrino Physics},  arXiv:1208.0512 (2012);
T.R. Edgecock et al., {\em The EUROnu Project}, arXiv:1305.4067 (2013).
\bibitem{cao-old} L. Zhan, Y. Wang, J. Cao and L. Wen, Phys. Rev. D79, 073007 (2009) [arXiv:0901.2976v2].
\bibitem{opera-5} N. Agafonova et al. (OPERA collaboration), Phys. Rev. Lett. 115, 121802 (2015) [arXiv:1507.01417].
\bibitem{palazzo-lbl} A. Palazzo, Phys. Rev. D 91 091301(R)(2015) [arXiv:1503.03966v3].
\bibitem{stanco-lbl} L. Stanco (for the OPERA collaboration), talk given at EPS 2015, Vienna, Austria, July 22--29, 2015, and arXiv:15010.04151 (2015).
\bibitem{giunti-book} C. Giunti and C. W. Kim, {\em Fundamentals of Neutrino Physics and Astrophysics}, Oxford University press (2007).
\bibitem{gg} G. Giacomelli, M. Giorgini,  L. Patrizii and M. Sioli, Adv. High Ener. Phys., 464926 (2013).
\bibitem{king} S. F. King, A. Merle, S. Morisi, Y. Shimizu and M. Tanimoto, New Jour. Phys. 16, 045018 (2014) [arXiv:1402.4271].
\bibitem{verma} S. Verma, Adv. High Ener. Phys., 385968 (2015).

\bibitem{fogli2011} G. L. Fogli, E. Lisi, A. Marrone, A. Palazzo and A. M. Rotunno, Phys. Rev. D 84, 053007 (2011) [arXiv:1106.6028].
\bibitem{dayabay} J. K. Ahn et el. (RENO collaboration), Phys. Rev. Lett.108, 191802 (2012) [arXiv:1204.0626];
F. P. An et al. (Daya Bay collaboration), Phys. Rev. Lett. 108, 171803 (2012) [arXiv:1203.1669].
\bibitem{theta13} F. P. Ahn et al. (Daya Bay collaboration), 	Phys. Rev. Lett. 112, 061801 (2014) [arXiv:1310.6732];
Y. Abe et al. (Double Chooz collaboration), J. High Energy Phys. 10, 086 (2014); Erratum ibid. 02, 074 (2015) [arXiv:1406.7763v4].
\bibitem{wolfe} L. Wolfenstein, Phys. Rev. D 17, 2369 (1978).
\bibitem{sw} S. P. Mikheyev and A. Y. Smirnov, Sov. J. Nucl. Phys., 42, 913 (1985).
\bibitem{macro} M. Ambrosio et al. (MACRO collaboration), Phys.Lett.B 434, 451 (1998) [arXiv:hep-ex/9807005].
\bibitem{k2k-2006} M. H. Ahn et al. (K2K collaboration), Phys.Rev.D74, 072003 (2006) [arXiv:hep-ex/0606032v3].
\bibitem{minos-2008} P. Adamson et al. (MINOS collaboration), Phys.Rev.Lett.101, 131802 (2008) [arXiv:0806.2237].
\bibitem{minos-2013} P. Adamson et al. (MINOS collaboration), Phys. Rev. Lett. 110, 251801 (2013) [arXiv:1304.6335v3].
\bibitem{t2k-2013} K. Abe et al. (T2K collaboration), Phys. Rev. Lett. 111, 211803 (2013) [arXiv:1308.0465v2].
\bibitem{T2K-2015-theta23} K. Abe et al. (T2K collaboration), Phys. Rev. D 91, 072010 (2015) [arXiv:1502.01550v2]. 
\bibitem{minos-2014} P. Adamson et al. (MINOS collaboration), Phys. Rev. Lett. 112, 191801 (2014) [arXiv:1403.0867v2].
\bibitem{icarus-sterile} M. Antonello et al. (ICARUS collaboration), Eur. Phys. J C73, 2599 (2013) [arXiv:1209.0122v4].
\bibitem{opera-sterile} N. Agafonova et al. (OPERA collaboration), JHEP 07, 004 (2013), addendum JHEP 07, 085 (2013)
[arXiv:1303.3953v2].
\bibitem{lisi-2015} F. Capozzi, G.L. Fogli, E. Lisi, A. Marrone, D. Montanino and A. Palazzo, Phys. Rev. D 89, 093018 (2014) [arXiv:1312.2878v2].
\bibitem{valle-2014} D. Forero, M. Tortola, and J. Valle, Phys.Rev. D90, 093006 (2014) [arXiv:1405.7540v3].
\bibitem{schwetz-deltacp} J. Elevant and T. Schwetz, arXiv:1506.07685v2 (2015).
\bibitem{palazzo-2015} A. Palazzo, arXiv:1509.03148 (2015).
\bibitem{lsnd} A. Aguilar et al. (LSND collaboration), Phys. Rev. D 64, no. 11, 112007 (2001) [arXiv:hep-ex/0104049v3].
\bibitem{lsnd-1995} C. Athanassopoulos et al. (LSND collaboration), Phys. Rev. Lett. 75, 2650 (1995).
\bibitem{giunti-1999} S.M. Bilenky, C. Giunti and W. Grimus, Prog. Part. Nucl. Phys. 43, 1 (1999) [arXiv:hep-ph/9812360v4].
\bibitem{mini-mu} A. A. Aguilar--Arevalo et al. (MiniBooNE collaboration), Phys.\ Rev.\ Lett. 103, 061802 (2009) [arXiv:0903.2465].
\bibitem{mini-sci-mu} K.~B.~M.~Mahn et al. (MiniBooNE and SciBooNE collaborations), Phys.\ Rev.\ D 85, 032007 (2012) [arXiv:1106.5685].
G. Cheng {\em et al.}, Phys.\ Rev.\ D 86, 052009 (2012) [arXiv:1208.0322].
\bibitem{reactor-sterile} G.  Mention et al.,  Phys.Rev. D 83, 073006 (2011)  [arXiv:1101.2755], and references therein.
D.~Lhuillier, Nucl.\ Phys.\ B, Proc.\ Suppl. 235-236, 11 (2013);
T.~Lasserre, Nucl.\ Phys.\ Proc.\ Suppl. 235-236, 214 (2013) [arXiv:1209.5090].
\bibitem{source} J. N. Abdurashitov et al. (SAGE collaboration), Phys. Rev. C 80, 015807 (2009);
J. N. Abdurashitov et al. (SAGE collaboration), Phys. Rev. C 73, 045805 (2006);
F. Kaether, W. Hampel, G. Heusser, J. Kiko, and T. Kirsten, Phys. Lett. B 685, 47 (2010); 
W. Hampel et al. (GALLEX collaboration), Phys. Lett. B 420, 114 (1998).
\bibitem{whitep} K. Abazajian et al., {\em Light Sterile Neutrinos: A White Paper}, arXiv:1204.5379 (2012).
\bibitem{fermi-sbl}  R. Acciarri et al. (ICARUS, LAr1-ND, MicroBooNE collaborations), (2015), arXiv:1503.01520.
\bibitem{reactor-new} T. Lasserre, talk given at NeuTel2015, Venice (Italy), 2-6 March 2015.
\bibitem{source-new} C. Lane et al. (NuLat collaboration),  arXiv:1501.06935 (2015);
P.  del Amo Sánchez (STEREO collaboration), talk given at EPS 2015, Vienna, 22-29 July 2015.
\bibitem{nessie-prop} A. Anokhina et al. (NESSiE collaboration), arXiv:1503.07471v2 (2015).
\bibitem{stanco-muapp} L. Stanco, S. Dusini, A. Longhin, A. Bertolin and M. Laveder, Ad.\ High\ Energy\ Phys., 948626 (2013) [arXiv:1306.3455v2].
\bibitem{icecube} M. G. Aartsen et al. (IceCube collaboration), Phys. Rev. D89, 062007 (2014) [arXiv:1311.7048];
see also the critical discussion in M. Lindner, W. Rodejohann, Xun-Jie Xu,  arXiv:1510.00666 (2015).
\bibitem{minosp} A. Timmons (MINOS collaboration), in proceedings of NuPhys2014, London 15--17 December 2014 (UK), arXiv:1504.04046 (2015);
see also Section 6 of T. Nakaya and R. K. Plunkett, arXiv:1507.08134v3 (2015).

\bibitem{mh-qian} X. Qian et al., Phys.Rev. D86, 113011 (2012) [arXiv:1210.3651].
\bibitem{mh-ciufoli} Emilio Ciuffoli, Jarah Evslin and Xinmin Zhang, JHEP 01, 95 (2014) [arXiv:1305.5150].
\bibitem{mh-schwetz} M. Blennow, P. Coloma, P. Huber and T. Schwetz, JHEP 1403, 028 (2014) [arXiv:1311.1822].
\bibitem{mh-blennow-bayes} M. Blennow, JHEP 01, 139 (2014) [arXiv:1311.3183].
\bibitem{wilks} S. S. Wilks, The Annals of Mathematical Statistics 9, 60 (1938).
\bibitem{mh-sk} Y. Hayato, talk at the 25th International Workshop on Weak Interactions and Neutrinos (WIN2015),
June 8--13, 2015, MPIK Heidelberg, Germany.
\bibitem{cowan} G. Cowan, K. Cranmer, E. Gross and O. Vitells, EPJC 71, 1554 (2011) [arXiv:1007.1727v3].
\bibitem{stat-higgs} ATLAS and CMS collaborations, ATLAS-CONF-2011-157, CMS-PAS-HIG-11-023 (2011), http://cdsweb.cern.ch/record/1399599.
\bibitem{higgs} The ATLAS collaboration, Phys.Lett. B 716, 1 (2012)  [arXiv:1207.7214v2];
The CMS collaboration, Phys. Lett. B 716, 30 (2012) [arXiv:1207.7235v2].
\bibitem{read} A. L. Read, Jour. Phys. G 28, 2693 (2002), and references [3], [9] therein.
\bibitem{nessie} A. Anokhina et al. (NESSiE collaboration), FNAL--P--1057, arXiv:1404.2521 (2014).
\bibitem{daya-bay-sterile} F. P. An et al. (Daya Bay collaboration), Phys. Rev. Lett. 113, 141802 (2014) [arXiv:1407.7259v2].
\bibitem{mh-lisi-1} F. Capozzi, E. Lisi and A. Marrone, Phys. Rev. D 89, 013001 (2014) [arXiv:1309.1638].
\bibitem{mh-japon} S.-F.Ge, K. Hagiwara, N. Okamura and Y. Takaesu, JHEP 1305, 131 (2013) [arXiv:1210.8141v2].
\bibitem{mh-lisi-2} F. Capozzi, E. Lisi and A. Marrone, arXiv:1508.01392 (2015).
\bibitem{nova1} R. Patterson (NOvA collaboration), talk given at the Joint Experimental-Theoretical Physics Seminar, Fermilab, 6 August 2015.
\bibitem{nova2} P. Adamson et al. (NOvA collaboration), submitted to Phys. Rev. Lett., arXiv:1601.05022 (2016).
\bibitem{Fel-Cou} G. J. Feldman and R. D. Cousins, Phys.\ Rev.\ D 57, 3873 (1998) [arXiv:physics/9711021v2].
\bibitem{mh-stanco} L. Stanco et al., {\em The determination of the Mass Hierarchy in the near future}, paper in preparation.
\bibitem{t2k-prop} Letter of intent: Neutrino oscillation experiment at JHF (2003),\\
http://neutrino.kek.jp/jhfnu/loi/loi\_JHFcor.pdf.
\bibitem{T2K-theta13-2014} K. Abe et al. (T2K collaboration), Phys. Rev. Lett., 112, 061802 (2014) [arXiv:1311.4750].
\bibitem{T2K-ptep} K. Abe et al. (T2K collaboration), Prog. Theor. Exp. Phys., 043C01 (2015) [arXiv:1412.0194].
\bibitem{T2K-antinue} M. R. Salzgeber (T2K collaboration), talk given at EPS 2015, Vienna, 22--29 July 2015, and 
arXiv:1508.06153 (2015).
\bibitem{nova-exp} D. S. Ayres et al. (NOvA collaboration), FERMILAB-DESIGN-2007-01 (2007);
 R. B. Patterson (NOvA collaboration), Nucl. Phys. Proc. Suppl. 235-236, 151 (2013).
 \bibitem{juno-cdr} JUNO collaboration, {\em JUNO Conceptual Design Report}, arXiv:1508.07166 (2015).
 \bibitem{juno-physics} JUNO collaboration, {\em Neutrino Physics with JUNO}, arXiv:1507.05613 (2015).
\bibitem{mh-may2015} X. Qian and P. Vogel, arXiv:1505.01891v3 (2015).

\bibitem{leps} Nakano et al. (LEPS collaboration), Phys. Rev. C 79, 025210 (2009) [arXiv:0812.1035].
\bibitem{lhcb} R. Aaij et al. (LHCb collaboration), Phys. Rev. Lett. 115, 072001 (2015) [ arXiv:1507.03414v2].
\end{thebibliography}
\end{document}